\documentclass[twocolumn,prd,aps,epfs,epsfig]{revtex4-1}

\usepackage{graphicx}%
\usepackage{dcolumn}
\usepackage{amsmath,amsthm,amssymb}
\usepackage[utf8]{inputenc}
\usepackage{orcidlink}

\begin{document}

\title{Similarity in the Arnowitt-Deser-Misner Mass from  Brill and Teukolsky Initial Data Sets Beyond the Linear Approximation}

%\author{M. A. Alcoforado$^{1}$, W O Barreto$^{1, 2}$ and H. P. de Oliveira$^1$}
%\address{$^2$ Centro de F\'{\i}sica Fundamental, Universidad de Los Andes, M\'erida 5101,  Venezuela}
%\address{$^1$ Departamento de F\'{\i}sica Te\'orica - Instituto de F\'{\i}sica
%	A. D. Tavares, Universidade do Estado do Rio de Janeiro, 
%	R. S\~ao Francisco Xavier, 524. Rio de Janeiro, RJ, 20550-013, Brazil}

\author{R. F. Aranha \orcidlink{0000-0002-9832-2379}}%
 \email{rafael.aranha@uerj.br}
\author{H. P. de Oliveira \orcidlink{0000-0003-1996-5016}}%
 \email{henrique.oliveira@uerj.br} 
\affiliation{%
Department of Theoretical Physics - Institute of Physics\\
Rio de Janeiro State University - 524 São Francisco Xavier Street - 20550-900\\ 
Maracanã - Rio de Janeiro - RJ - Brazil\\  
}%


\date{\today}
% \author{R. F. Aranha}
% \email{rafael.aranha@uerj.br}
% \affiliation{
% Departamento de F\'{\i}sica Te\'orica - Instituto de F\'{\i}sica
% A. D. Tavares, Universidade do Estado do Rio de Janeiro \\ 
% R. S\~ao Francisco Xavier, 524. Rio de Janeiro, RJ, 20550-013, Brazil}

% \author{H. P. de Oliveira}\thanks{Corresponding author}
% \email{henrique.oliveira@uerj.br}
% \affiliation{
% Departamento de F\'{\i}sica Te\'orica - Instituto de F\'{\i}sica
% A. D. Tavares, Universidade do Estado do Rio de Janeiro \\ 
% R. S\~ao Francisco Xavier, 524. Rio de Janeiro, RJ, 20550-013, Brazil}

% \date{\today}

\begin{abstract}
The two most common initial data for vacuum axisymmetric spacetimes are the Brill and Teukolsky gravitational waves. The subsequent numerical evolution of these data exhibits distinct properties mainly associated with the critical gravitational collapse. A possible way of understanding these differences is first to look at the linear level, where the Brill waves have a multipolar structure, while the Teukolsky waves have a quadrupolar structure. Despite being structurally distinct at the linear and nonlinear levels, we show that these gravitational wave initial data share an unexpected similarity related to the distribution of ADM mass as a function of the initial wave's amplitude. More specifically, both configurations satisfy a nonextensive relation, commonly seen in systems governed by long range interactions.
%\vspace{8cm}
%“Essay written for the Gravity Research Foundation 2024 Awards for Essays on %Gravitation.”
\end{abstract}

\maketitle

The direct detection of gravitational waves by the LIGO consortium in 2015 \cite{LIGO} rewarded decades of experimental and theoretical efforts to this end since the pioneering attempts of Weber in the 60's \cite{weber}. It represented the confirmation of one of the most spectacular predictions of the General Theory of Relativity (GTR) in which the perturbations or ripples of spacetime propagate with the speed of light in vacuum.  %We mention that the indirect evidence on the existence of gravitational waves is provided by the observations of the Hulse-Taylor binary stellar system \cite{HT}. 
The direct detection of the gravitational radiation opened a new window to observe the Universe with the potential of revealing physical aspects of the very early Universe, the interaction of binary black holes, black holes and neutron stars, or even the inner structure of neutron stars. %We can say that we have witnessed the birth of gravitational wave astronomy. The lesson is that aspects of the source leave fingerprints in the observed templates of the received waves.
Aside from the promising observational of exploring the universe with gravitational waves, there are theoretically relevant issues about the dynamics of gravitational waves.

The collapse of pure gravitational waves in connection with critical phenomena is a formidable theoretical problem under current scrutiny. We recall that Choptuik \cite{choptuik} first reported the critical phenomena in the gravitational collapse in connection with the dynamics of spherically symmetric massless scalar fields. The observed properties resemble to the critical behavior in another fields, more notably when the evolution of one-parameter initial data is close to the critical parameter, say $\eta_*$. In this regime and considering supercritical solutions with $\eta \sim \eta_*$, the mass of the newly formed black holes satisfies an approximate power-law, $M \propto (\eta-\eta_*)^\gamma$, where $\gamma$ depends on the matter field under consideration but not on the particular initial data family. Moreover, when the parameter is adjusted close to its critical value, the dynamics exhibit a discrete or continuous self-similarity phase depending on the matter model.

When dealing with the critical phenomena in gravitational collapse beyond the spherically symmetric case, the most simple and intriguing situation is the collapse of vacuum gravitational waves \cite{ab_ev1,ab_ev2,hild_13,hild_17,led_khir_21}. In general, two distinct initial data are considered to feed the Einstein field equations: the Brill waves \cite{brill} and Teukolsky waves \cite{teukolsky}. The first represents nonlinear gravitational waves while the latter are exact quadrupolar solutions of the linearized field equations (description of more details about the current studies on the dynamics of vacuum gravitational waves).

Recently, Fernandez et al. \cite{fernandez}, motivated by numerical studies of critical phenomena in the gravitational collapse of vacuum gravitational waves, tried to understand the peculiarities and differences when initial data sets constituted by Brill and Teukolsky waves are evolved. They have compared these initial data sets at the linear regime. Still, surprisingly, after considering Gaussian seed functions for both Brill \cite{eppley,holz} and Teukolsky waves \cite{baumg_shapiro}, they concluded that their initial data are similar. This unexpected result arises from the distinct nature of both initial data, i.e., while Teukolsky waves are manifestly quadrupolar, the Brill waves contain higher multipole moments.

We report here a remarkable aspect of Brill and Teukolsky's initial data sets. More specifically, we calculate the Arnowitt-Deser-Misner mass \cite{ADM} generated by these distinct initial gravitational wave distributions and show that it scales similarly with the initial amplitude parameter, $\epsilon$, present in both Brill and Teukolsky seed functions. We recall that the ADM mass is the total spacetime mass-energy content. Therefore, in the present case of pure gravitational waves, its value results from the contribution of all distinct multipole moments, not only in the pure quadrupolar Teukolsky waves but also in the whole multipolar spectrum of the Brill waves. 

We briefly describe the determination of the ADM formalism for the axisymmetric Brill and Teukolsky initial data sets. Following the $3+1$ initial value problem for Einstein's field equations \cite{baumg_shapiro,cook}, we start with the three-dimensional line element expressed as

\begin{equation}
	dl^2=\Psi^4\,d\bar{l}^2,%\Psi^4\,\bar{\gamma}_{ij} dx^i dx^j, 
\end{equation} 

\noindent where $\Psi$ is the conformal factor, $d\bar{l}^2=\bar{\gamma}_{ij} dx^i dx^j$, and the metric components $\bar{\gamma}_{ij}$ are specified by the choices of Brill or Teukolsky initial data sets. We choose time-symmetric initial data characterized by the vanishing of the extrinsic curvature, $K_{ij}=0$, implying that the momentum constraint is identically satisfied. We determine the conformal factor after solving the Hamiltonian constraint equation written for the spatial metric (1) as

% \begin{equation}
% R+K^2-K_{ij}K^{ij} = 0 \Rightarrow R = \bar{\nabla}^2\Psi-\frac{1}{8} \bar{R}\Psi = 0,
% \end{equation}

\begin{equation}
R = \bar{\nabla}^2\Psi-\frac{1}{8} \bar{R}\Psi = 0,
\end{equation}

\noindent where $R$ is the Ricci scalar associated with the line element (1), $\bar{\nabla}$ and $\bar{R}$ are the Laplacian operator and the Ricci scalar due to the metric components $\bar{\gamma}_{ij}$, respectively. Therefore, the Hamiltonian constraint (2) is transformed into an elliptic equation for the conformal factor $\Psi$ whose solution determines uniquely the initial data for the problem.%, say, the initial metric components $\gamma_{ij}$. 

The ADM mass measures the total mass-energy of an asymptotically flat spacetime. It is a conserved quantity, whose definition consists of a two-dimensional surface integral at infinity \cite{baumg_shapiro} taken into account the initial metric components $\gamma_{ij}$ which are determined once the Hamiltonian constraint is solved. 

The three-dimensional line element for Brill waves is \cite{brill}

\begin{equation}
dl^2 = \Psi^4\,\left[\mathrm{e}^{2q} (d\rho^2+dz^2) + \rho^2 d\phi^2\right],
\end{equation}

\noindent where $(\rho,\phi,z)$ are the cylindrical coordinates and $q=q(\rho,z)$ is the seed function. The two most popular choices for the seed function that satisfy the appropriate boundary conditions are \cite{eppley,holz}

\begin{eqnarray} 
	q(\rho,z) &=& \epsilon\rho^2\mathrm{e}^{-(\rho^2+z^2)/\sigma^2},\\
	\nonumber \\
	q(\rho,z) &=& %\frac{\epsilon\rho^2}{1+\left(\frac{\sqrt{\rho^2+z^2}}{\sigma}\right)^{n/2}},
	\epsilon\frac{\rho^2}{1+\left(\frac{\rho^2+z^2}{\sigma^2}\right)^{n/2}},\;\;n \geq 4,
\end{eqnarray} 

\noindent known as Holz and Eppley's seed functions, respectively. The parameters $\sigma$ and $\epsilon$ are related respectively to the width and the initial amplitude of the Brill gravitational wave. The ADM mass is calculated through the surface integral

\begin{equation}
M_{ADM}  = \int_{-\infty}^\infty\,\int_{0}^\infty\,\frac{1}{\Psi^2}\,\left[\left(\frac{\partial  \Psi}{\partial \rho}\right)^2+\left(\frac{\partial  \Psi}{\partial z}\right)^2\right] \rho d\rho dz,
\end{equation}

\noindent where the conformal factor is determined once the Hamiltonian constraint (2) is solved for the line element (3) considering the seed functions (4) and (5). We have set $\sigma=1$ and $n=4$.

For the the Teukolsky initial data, we have the following three-dimensional line element
\begin{eqnarray}
dl^2 &=& \Psi^4\,\big[h_{11}\,dr^2+2h_{12}\,rdrd\theta+r^2(h_{22}\,d\theta^2 \nonumber\\
\nonumber \\
&&+h_{33}\,\sin^2\theta\,d\phi^2)\big],
\end{eqnarray}

\noindent where $(r,\theta,\phi)$ are the spherical coordinates. The metric coefficients $h_{ij}=h_{ij}(r,\theta)$ are the exact linear solution of Einstein's equations describing quadrupolar gravitational waves \cite{teukolsky}. These metric coefficients can be constructed from the seed function $F(t,r)=F_1(t-r)+F_2(t+r)$, where $F_1$ and $F_2$ refer to incoming and outgoing solutions respectively. We adopt the choice of Hilditch et al. \cite{hild_13}, $F_1=-F2$, resulting in a solution at $t=0$ with moment of time symmetry ($K_{ij}=0$) and 

\begin{equation}
F_1(t-r)=\epsilon\frac{(t-r)}{2\sigma}\left[\mathrm{e}^{-[t-(r+r_0)]^2/\sigma^2}+ \mathrm{e}^{-[t-(r-r_0)]^2/\sigma^2} \right].
\end{equation}

\noindent Here $\epsilon$ is the wave amplitude, $\sigma$ its width, and $r_0$ locates the center of the wave. We determine the ADM mass after solving the integral

\begin{equation}
M_{ADM}=-\lim_{r \rightarrow \infty}\,\int_{0}^{\pi}\,\left(\frac{\partial \Psi}{\partial r}r^2\right)\sin \theta\,d\theta,
\end{equation}

\noindent where the conformal factor $\Psi$ is obtained once again by solving the Hamiltonian constraint (2) and taking into account the exact expression for $h_{ij}$, which is generated by the seed function (8) with $\sigma=1/2$ and $r_0=2$ \cite{hild_13}.

We have solved the Hamiltonian constraint (2) numerically for both Brill and Teukolsky initial data using a spectral method developed in Ref. \cite{aranha_hpo}, with the conformal factor expressed by
\begin{equation}
\Psi = 1 + \delta \Psi,
\end{equation}
\noindent where $\delta \Psi$ \textit{is not} an approximation. %We calculate the ADM mass for these gravitational wave initial data considering $\epsilon > 0$ ranging from tiny values characterizing the linearized gravitational wave regime up to higher values producing a nonlinear evolution.
We calculate the ADM mass for these gravitational wave initial data considering $\epsilon > 0$ ranging from very small values characterizing the linearized gravitational wave regime up to higher values which are able to produce nonlinear evolutions.

\begin{figure}[htb]
\includegraphics[width=7.5cm,height=6.5cm]{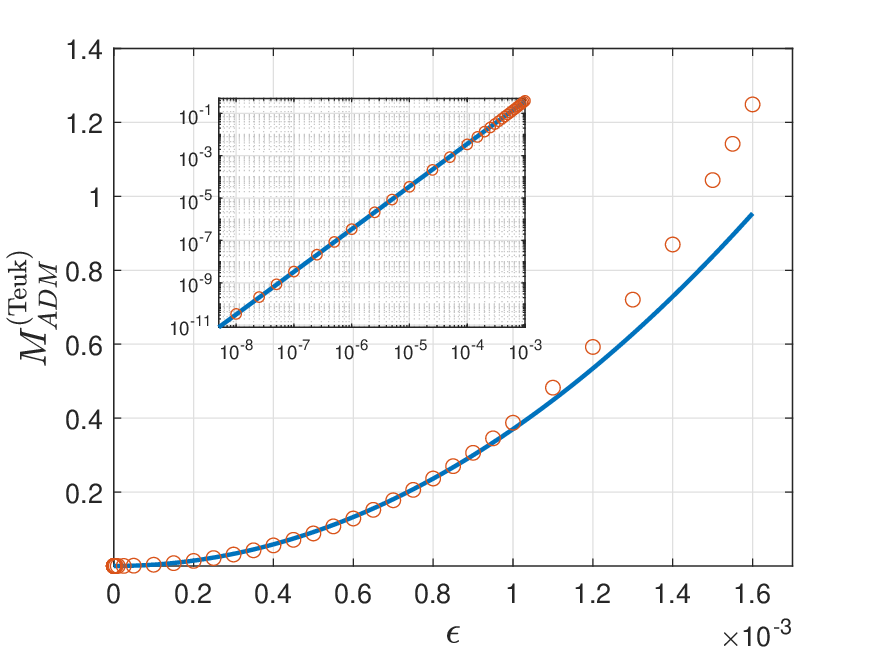}
%\center{(a)}
\includegraphics[width=7.5cm,height=6.5cm]{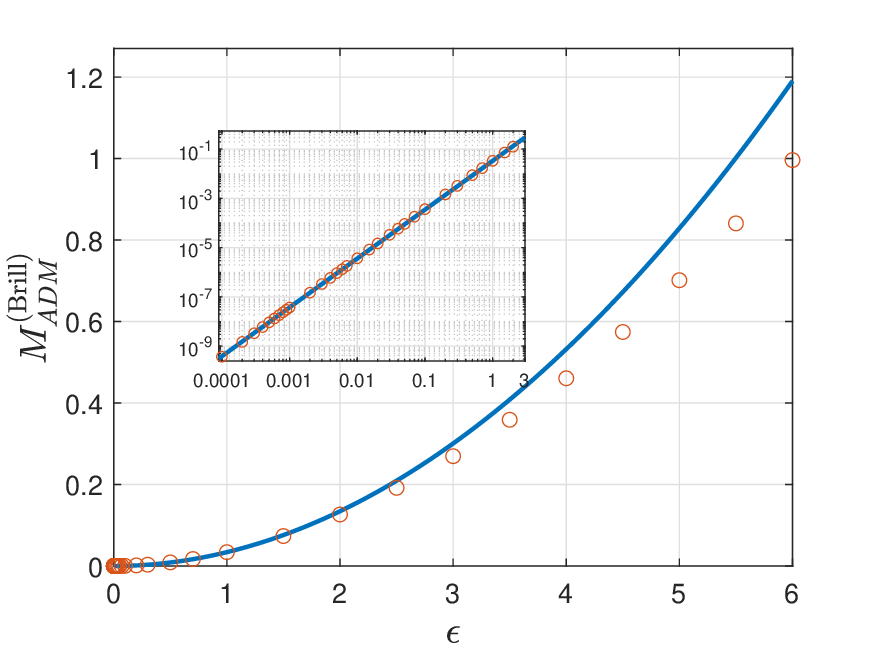}
\caption{We present plots of the scaling law $M_{ADM} \propto \epsilon^2$ (continuous line) for the Teukolsky (upper panel) and Brill (lower panel) initial data, where circles represent the numerically generated data. We have used the Holz seed function for the Brill waves (cf. Eq.(4)). Notice that the quadratic scaling law holds beyond $\epsilon \ll 1$, but eventually, for higher values, it fails. We display the insets with a log-log scale plot in each case in the regions such that the quadratic scaling law holds.}%of the ADM mass, the quadratic scaling fails. We display the insets with a log-log scale plot in each case in the regions such that the quadratic scaling law holds.}
\end{figure}

%Let us focus on the smaller values of $\epsilon \ll 1$. Thus, the Hamiltonian constraint (2) for the Brill waves dictates that $\delta \Psi \sim \mathcal{O}(\epsilon)$, no matter the adopted seed function.
%Let us focus on the smaller values of $\epsilon \ll 1$. As a consequence, the Hamiltonian constraint (2) corresponding to the Brill waves dictates that $\delta \Psi \sim \mathcal{O}(\epsilon)$, no matter the adopted seed function. Then, Eq. (6) for the ADM mass demands that

For some analytical considerations, let us focus first on the smaller values of the initial amplitude, $\epsilon \ll 1$. Inspecting the Hamiltonian constraint (2) for the Brill waves (see Eq. (40) of Ref. \cite{aranha_hpo}), one can show that $R \sim \mathcal{O}(\epsilon)$ which yields $\delta \Psi \sim \mathcal{O}(\epsilon)$, no matter the adopted seed function (4) or (5). Consequently, Eq. (6) demands that 
\begin{equation}
M_{ADM}^{\mathrm{(Brill)}} =   \mathcal{O}(\epsilon^2).
\end{equation}   

%\noindent Considering once again the Hamiltonian constraint (2) but now for the Teukolsky waves, we conclude that, distinctly from the previous case,  $\delta \Psi \sim \mathcal{O}(\epsilon^2)$. In this manner, Eq. (9) shows that
\noindent Now considering the Hamiltonian constraint (2) for the Teukolsky waves (see Eq. (35) of Ref. \cite{aranha_hpo}), we conclude that, distinctly from the previous case, $R \sim \mathcal{O}(\epsilon^2)$ implying, as a consequence  
$\delta \Psi \sim \mathcal{O}(\epsilon^2)$. In this manner, from Eq. (9), it follows that the ADM mass scales with $\epsilon$ as

\begin{equation}
M_{ADM}^{\mathrm{(Teuk)}} =   \mathcal{O}(\epsilon^2),
\end{equation}  

%\noindent Despite the conformal factor scales distinctly for each Brill and Teukolsky initial data, both ADM masses are proportional to the square of the initial gravitational wave amplitude.
\noindent Although the conformal factor scales distinctly for each Brill and Teukolsky initial data, both ADM masses are proportional to the square of the initial gravitational wave amplitude.

%Remarkably, by increasing the initial amplitude $\epsilon$ beyond the linear approximation, we notice that the scale relations given by Eqs. (11) and (12) hold. 
The numerical values of the ADM masses of both Brill and Teukolsky initial data agree with the analytical estimates for $\epsilon \ll 1$. Remarkably, by increasing $\epsilon$ beyond the linear approximation, say for $\epsilon \sim \mathcal{O}(1)$, the scale relations (11) and (12) are still valid. However, continuing to increase $\epsilon$ makes these scaling relations no longer valid. We summarized both results in Fig. 1, showing the validity of the scaling laws Eqs. (11) and (12) (inset plots) and the complete numerical data represented by circles for both Brill and Teukolsky initial data, in the former case taking into account the Holz seed function. The critical amplitude found in numerical studies are $\epsilon \gtrsim 1.5 \times 10^{-3}$ and $\epsilon \gtrsim 4.6$, for the Teukolsky and Brill data, respectively, with the chosen seed functions \cite{hild_13}.

%The immediate question is whether there is an identical scaling law for the ADM mass considering both Brill and Teukolsky initial data, such that these also fit the data resulting from higher initial amplitudes. The answer is positive and the scaling law is given by:
%
The inspection of the plots of Fig. 1 raises the question of whether there is a unique scaling law for the initial ADM masses for the Brill and Teukolsky initial data covering from small to higher amplitudes $\epsilon$. Inspired by the studies of nonextensive statistics \cite{tsallis}, we found that  

\begin{equation}   
M_{ADM} \propto \epsilon^\gamma\,(1+\epsilon^\gamma)^\alpha,
\end{equation}   

%\noindent where we obtained $\gamma \approx 2$, and $\alpha$ has distinct values depending on whether we are using Brill or Teukolsky waves as the initial data. According to the plots of Fig. 2, the agreement between the numerical data and the scaling law (13) attests to its validity.

\noindent where $\gamma \approx 2$ no matter is the initial data, and $\alpha$ assumes distinct values depending on the initial data under consideration. It becomes clear that for very small values of $\epsilon \ll 1$, the scaling law (13) reproduces correctly the analytical estimate given by Eqs. (11) or (12), the whole numerical data is reproduced with the correct choices for the parameter $\alpha$. The best fit indicates that the parameter $\alpha$ depends on the initial data according to

\begin{eqnarray}
\alpha_{\mathrm{(Brill)}} &=& \begin{cases}
    &-0.1673\quad \text{Eppley seed function} \\
    \nonumber \\
    &-0.0565\quad \text{Holz seed function}
  \end{cases} \\
 \nonumber \\
\alpha_{\mathrm{(Teuk)}} &=& 1.617 \times 10^5 \nonumber
\end{eqnarray}

\begin{figure}[htb]
\includegraphics[width=7.5cm,height=6.5cm]{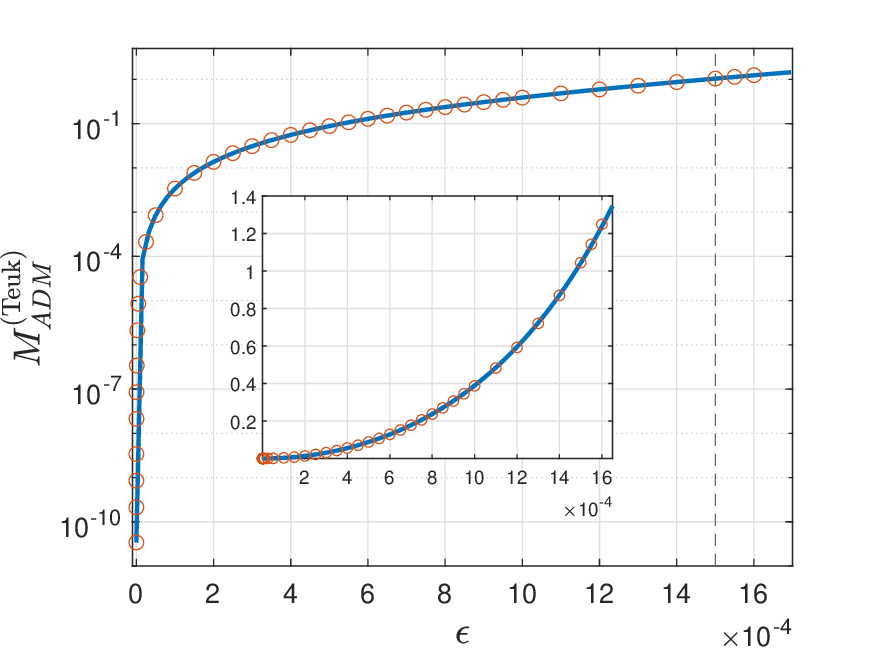}
\includegraphics[width=7.5cm,height=6.5cm]{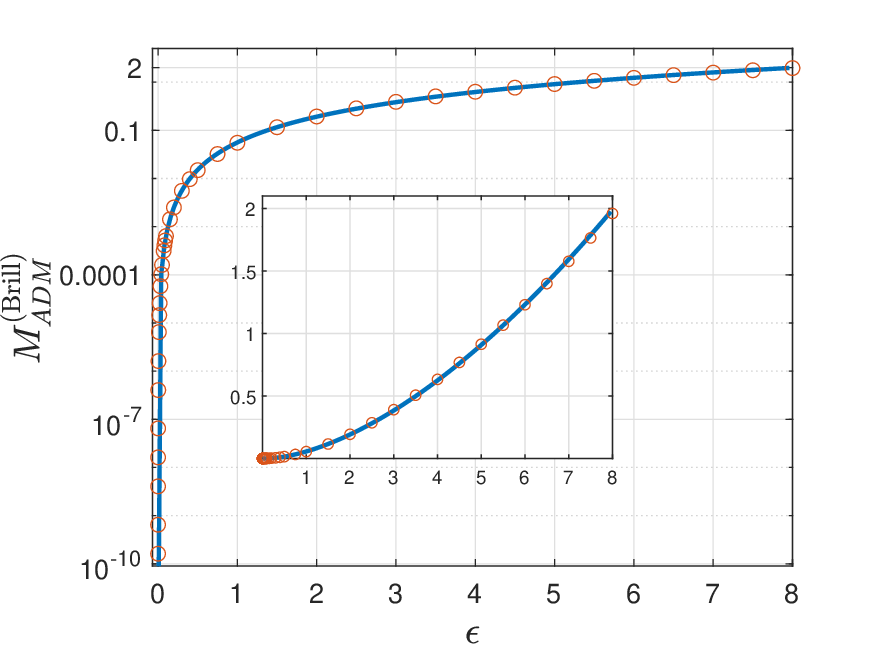}
\caption{Log-linear plots of the $M_{ADM}$ masses using the nonextensive scaling law (13) (continuous line) for Teukolsky and Brill's initial data with the Eppley's seed function (cf. Eq. (5)) depicted by the upper and lower panels, respectively. Circles represent the numerical data. In both cases, $\gamma \approx 2$, but $\alpha$ is distinct, namely $\alpha_{\mathrm{(Teuk)}} \simeq 1.617\times 10^5$ and $\alpha_{\mathrm{(Brill)}} \simeq -0.1673$. Still, for the Brill initial data with the Holz seed function, $\alpha_{\mathrm{(Brill)}} \simeq -0.0565$. We show the linear scale plots with scaling law (13) in the insets. For reference, $\epsilon  \gtrsim 1.5 \times 10^{-3}$ signalizes the critical amplitude for the Teukolsky initial data \cite{hild_13}.}
\end{figure}

\noindent We show in Fig. 2 the agreement of the scaling law (13) and the numerical data for the Teukolsky and the Brill initial data. The panels display the logarithmic scale plot and the insets, the linear scale plot. 

%In our final remarks, we point out that despite the distinct structure of the initial data representing Brill and Teukolsky gravitational waves, the ADM mass satisfies the scaling law (13) for both cases. This expression is inspired by the nonextensive statistical distribution \cite{tsallis}, specifically the called q-Weilbul distribution \cite{qdistributions}. We have already explored the role of nonextensive distributions related to the emission of gravitational waves from black holes, although in a simplified scenario \cite{deol_hon_ment_2008, prd_2008,ijmpd_2008}. It becomes evident that the long-range character of the gravitational interaction is the factor behind the validity of the distribution given by Eq. (13) for distinct gravitational wave initial data considering both linear and nonlinear regimes. The following steps instigated by this investigation are to consider the scaling of the ADM mass for negative initial amplitudes, $\epsilon < 0$, and, as a more advanced proposal, to seek the scaling relation associated with the rate of mass emitted in the formation of a black hole during the collapse of pure gravitational waves.     

In our final remarks, we point out that despite the distinct structure of the initial data representing Brill and Teukolsky gravitational waves, the ADM mass satisfies the same scaling law given by Eq. (13). This expression is inspired by the nonextensive statistical distribution \cite{tsallis}, more specifically the called q-Weilbul distribution \cite{qdistributions}. We have already explored the role of nonextensive distributions related to the emission of gravitational waves from black holes, although in a simplified scenario \cite{deol_hon_ment_2008, prd_2008,ijmpd_2008}. It becomes evident that the long-range character of the gravitational interaction is the determinant factor behind the validity of the distribution given by Eq. (13). The following step instigated by this investigation is to seek the scaling relation associated with the rate of mass emitted in the formation of a black hole during the collapse of pure gravitational waves.

\begin{acknowledgements}
H. P. de Oliveira thanks Conselho Nacional de Desenvolvimento Cient\'ifico e Tecnol\'ogico (CNPq) and Funda\c c\~ao Carlos Chagas Filho de Amparo \`a Pesquisa do Estado do Rio de Janeiro (FAPERJ)
(Grant No. E-26/200.774/2023 Bolsas de Bancada de Projetos (BBP)). We are grateful to Prof. Tsallis for some suggestions on the manuscript.
\end{acknowledgements}

\end{document}